# Stress-Testing Structure-Aware Calibration of Malware Graph Neural Networks under Type Shift

Junru Zhu
*Independent Researcher*
Seattle, USA
junru.zhuu@gmail.com

Yixin Yang
*Independent Researcher*
New York, USA
vicki_yang@chiefgroup.com.hk

Xiaoqing Ding
*University of Chicago*
Chicago, USA
alexading@uchicago.edu

Ruoyu Qi
*Independent Researcher*
Charlotte, USA
dxfreedom94@gmail.com

***Abstract*—Post-hoc malware calibrators can condition confidence on graph structure, but their structural inputs may leave the support represented by validation data under malware-type shift. We study this risk on MalNet-Tiny by holding out each of four malware types across three seeds, freezing a graph isomorphism network, and fitting post-hoc mappings only on known-type data. On known-type tests, generic and combined structural calibrators have negative log likelihood (NLL) of 0.4341 and 0.4628. Their mean held-out-type gap widens to 0.1369, with a four-type 95% interval of [−0.0125, 0.2864] and improvement in one of four types. Eight label-independent nuisance variables produce a smaller mean NLL increase of 0.0415, but the paired combined-versus-nuisance interval includes zero. We then use calibration data to define a community-support guard: predictions outside the 95th percentile of calibration-standardized community displacement revert to the generic calibrator. The guard reduces combined NLL by 0.1133 and high-confidence errors from 43.25 to 33.25 per 400-sample cell. It also outperforms matched-rate random fallback on the original folds, while generic and uncalibrated probabilities retain lower held-out-type NLL. This exploratory recovery localizes harmful combined predictions; the stress test exposes heterogeneous calibration failures without establishing a general structural-calibration penalty or a deployment gain over simpler alternatives.**

## I. Introduction

Graph neural networks (GNNs) classify malware from function-call structure [1–2], with public benchmarks such as MalNet supporting reproducible evaluation [3]. Confidence also determines which predictions receive automatic handling or further analysis. Post-hoc calibration fits probability mappings to validation data, with temperature scaling as a logit-only baseline [4].

Changes in malware types, families, collection pipelines, and prevalence can distort evaluation [5]. Studies of uncertainty, selective prediction, and longitudinal malware shift examine this risk [6–9]. A structural calibrator introduces another exposure: graph-level inputs can leave calibration support even when fitting partitions remain disjoint.

We test this exposure on MalNet-Tiny by excluding each of four malware types from training, early stopping, and calibration across three seeds. A frozen graph isomorphism network (GIN) supplies logits to matched eight-feature generic and community blocks. Proper scores, selective risk, known-type baselines, and support diagnostics measure their probability quality. Combined calibration has higher known-type negative log likelihood (NLL) than generic; the held-out-type gap widens to +0.1369, with a four-type interval that includes zero. A nuisance control leaves the mechanism unresolved. A calibration-only guard recovers combined loss and beats matched-rate random fallback on the original folds, while simpler alternatives retain lower held-out-type NLL.

Our contributions are:

- a leakage-controlled stress test in which four malware types are excluded from model training, early stopping, and calibration across three optimization seeds;
- a matched structural comparison that separates community covariates from generic graph properties and uses held-out type, rather than optimization seed, as the primary inference unit;
- known-type baselines and a capacity-matched nuisance-feature control with paired uncertainty estimates for the structural comparison;
- an exploratory label-free community-support guard, tested against random fallback with the same number of routed predictions.

## II. Related Work

### A. Graph-Based Malware Analysis

Malware analysis uses call, control-flow, application programming interface (API), and heterogeneous graphs [10]. Android and Windows detectors learn program structure [1–2]; MalNet supplies the 5,000-graph MalNet-Tiny benchmark [3]. The same graph-learning pattern extends beyond malware: Xiao et al. combine graph neural networks with multi-head attention to jointly correlate alarms and identify root alarms in optical transport networks [11]. We freeze the detector and test graph summaries in its post-hoc probability mapping.

*B. Confidence Under Malware Shift*

Malware studies use uncertainty for dataset shift and adversarial examples [6], uncertainty-aware detection [7], and selective-method stability [8]. CADE detects representation drift [12]; TESSERACT and LAMDA define realistic splits and longitudinal evaluation [5, 9]. We examine shifted inputs to a structural calibrator rather than changing the detector.

*C. Calibration Support and Structural Controls*

Covariate-shift calibration uses unlabeled target data for importance weighting or adaptation [13–14]. Graph methods use topology for node confidence [15], and structural shift analysis motivates label-free temperature adjustment [16]. Here whole-graph summaries enter a frozen malware calibrator. Their correlations with size, sparsity, and components motivate matched structural blocks, nuisance controls, and support-based fallback.

## III. Study Design

Each fold freezes a detector trained without the held-out type and fits calibrators on disjoint known-type data. Known-type tests, nuisance inputs, and matched-rate random fallback separate baseline fit from shifted recovery.

*A. Held-Out Malware-Type Protocol*

MalNet-Tiny contains 5,000 Android function-call graphs with official train, validation, and test splits [3]. We map benign to class zero and all malware types to class one. Each fold holds out addisplay, adware, downloader, or trojan from model training, early stopping, and calibration. Table I summarizes the fold protocol.

Known-type validation graphs are divided deterministically into equal early-stopping and calibration partitions. We retain three test views: known-type test graphs, 200 benign plus 200 held-out-malware graphs, and held-out malware alone. The balanced 400-sample view is the primary held-out-type test view. Because it has lower malware prevalence than the known-type test view, a diagnostic view combines all 200 held-out-malware graphs with 67 deterministically selected benign graphs.

Feature standardizers and calibrator parameters use calibration data only; test labels enter metric computation. Paired methods use identical graph identities in all 12 planned fold-seed cells, with 100,800 retained base predictions across methods and views.

TABLE I. Per-cell data allocation. Counts are identical across folds and seeds; the malware column includes only known types during fitting.

| Partition/view | Benign | Malware | Purpose |
|---|---|---|---|
| Train | 700 | 2100 | GNN fitting |
| Early stop | 50 | 150 | Checkpoint selection |
| Calibration | 50 | 150 | Post-hoc fitting |
| Known-type test | 200 | 600 | Reference view |
| Held-out-type test | 200 | 200 | Primary evaluation |
| Matched test | 67 | 200 | Prevalence check |
| Held-out only | 0 | 200 | Shift diagnosis |

*B. Detector and Training*

Directed call edges are retained for degree calculation, and message passing uses both directions. Each node receives five deterministic features: $\log(1+\text{in-degree})$, $\log(1+\text{out-degree})$, $\log(1+\text{total-degree})$, and source and sink indicators. The classifier is a three-layer GIN [17] with 64 hidden dimensions, global mean and max pooling, 0.2 dropout, and a binary logit head.

Training uses class-weighted binary cross entropy and Adam with batch size 32, learning rate $10^{-3}$, and weight decay $10^{-4}$. Early stopping minimizes known-type NLL with patience 12 and an 80-epoch cap. All runs use deterministic execution on a central processing unit (CPU), with seeds 7, 23, and 47.

*C. Post-Hoc Calibrators*

The six base estimators are the uncalibrated sigmoid, temperature scaling, Platt scaling, and logistic mappings from the logit plus generic, community, or both structural blocks. Fitted mappings use known-type calibration only.

The generic block contains log node count, log directed-edge count, density, mean and standard deviation of undirected degree, maximum-degree ratio, connected-component ratio, and transitivity. The community block uses Louvain partitions [18] on a simple undirected graph with self-loops removed and all nodes retained. NetworkX 3.6.1 uses seed 0, resolution 1, threshold $10^{-7}$, and no level cap; an edgeless graph has singleton communities and zero modularity. For $n$ nodes, $m$ undirected edges, $K$ communities, community sizes $n_k$, internal edge counts $m_k$, proportions $a_k = n_k/n$, and bridge-node set $B_{\text{bridge}}$, the eight features are

$$\boldsymbol{c} = (Q, \log(1+K), K/n, H/\log K, \max_k a_k, \sum_k m_k/m, |B_{\text{bridge}}|/n, \sum_k a_k \rho_k). \quad (1)$$

Here $Q$ is modularity, $H = -\sum_k a_k \log(a_k + 10^{-12})$, and $\rho_k = 2m_k/[n_k(n_k-1)]$ is zero for singleton communities. Entropy is zero for $K = 1$, and the intra-edge ratio is zero for $m = 0$. A bridge node touches an edge crossing communities. Platt and structural calibration use $L_2$-regularized logistic regression with $C = 1$ and at most 2,000 iterations. Logits and covariates are standardized from calibration data only.

Generic features measure scale, sparsity, degree heterogeneity, fragmentation, and clustering; community features measure partition organization. Equal dimensionality matches input count, while correlations and effective capacity can still differ. Combined tests the addition of community information after generic topology is observed.

Let $z_i$ denote the frozen GNN logit and let $\boldsymbol{g}_i, \boldsymbol{c}_i \in \mathbb{R}^8$ denote the generic and community blocks. Each logistic calibrator has the form

$$\hat{p}_i^{(m)} = \sigma\left(b_m + \boldsymbol{w}_m^{\mathrm{T}} \tilde{\boldsymbol{\phi}}_i^{(m)}\right), \quad (2)$$

where the tilde denotes calibration-fitted standardization. The design vectors are $z_i$ for Platt scaling, $[z_i; \boldsymbol{g}_i]$ for generic, $[z_i; \boldsymbol{c}_i]$ for community, and $[z_i; \boldsymbol{g}_i; \boldsymbol{c}_i]$ for combined. Generic and community therefore have nine inputs each, whereas combined has 17. Temperature scaling instead uses $\sigma(z_i/T)$ with one fitted scalar $T$.

We add two diagnostic methods without retraining the GNN. The capacity-matched control replaces the community block with eight deterministic standard-normal nuisance variables derived from graph identity. The variables are independent of labels and use the same generation rule in calibration and evaluation. We average metrics over 20 nuisance realizations. The community-support guard measures

$$s_i = \max_j \left| \frac{c_{i,j} - \mu_j^{\mathrm{cal}}}{\sigma_j^{\mathrm{cal}}} \right|, \tag{3}$$

and uses the generic probability when $s_i$ exceeds the 95th percentile of calibration scores; otherwise it retains the combined probability. Thresholds at calibration percentiles 90, 95, 97.5, and 99 use calibration covariates without target labels. A random-fallback control uniformly selects exactly the guard's fallback count within each fold-seed-view cell, without labels, over 200 deterministic repeats. We average its metrics before comparing methods; repeats do not add independent shifts.

### D. Metrics and Statistical Analysis

For binary label $y_i$ and malware probability $p_i$, NLL penalizes confident errors and Brier score measures squared probability error:

$$\mathrm{NLL} = -\frac{1}{n}\sum_i [y_i \log p_i + (1-y_i)\log(1-p_i)], \tag{4}$$

$$\mathrm{Brier} = \frac{1}{n}\sum_i (p_i - y_i)^2. \tag{5}$$

We also compute ten-bin fixed and adaptive expected calibration error (ECE), accuracy, macro F1, and area under the risk–coverage curve (AURC), which integrates risk as low-confidence predictions are rejected. The predeclared primary comparison is combined minus generic for NLL, Brier, and AURC. Capacity and guard analyses were added after the primary result and are treated as diagnostic comparisons. For each comparison, we average three seeds within each held-out type and use a 10,000-resample bootstrap over the four types [19]. A hierarchical diagnostic resamples both levels. Optimization seeds quantify within-type variability without being treated as independent malware-type shifts. ECE remains diagnostic because it depends on binning. Known-type intervals summarize the four folds, whose known-type test sets overlap; they do not represent four independent known-type populations.

For feature $j$ and held-out type $t$, standardized mean displacement is

$$d_{j,t} = \frac{\mu_{j,t}^{\mathrm{heldout}} - \mu_{j,t}^{\mathrm{cal}}}{\sigma_{j,t}^{\mathrm{cal}}}. \tag{6}$$

This support diagnostic is measured in calibration standard-deviation units. Graph-size analyses use four 100-sample node-count quartiles per cell. Intervals use bootstrap percentiles 2.5 and 97.5.

TABLE II. KNOWN-TYPE TEST PERFORMANCE, AVERAGED OVER FOUR FOLDS AND THREE SEEDS; 800 GRAPHS PER CELL. ECE USES TEN FIXED BINS.

| Method | NLL | Brier | ECE | AURC |
|---|---|---|---|---|
| Uncalibrated | 0.5030 | 0.1749 | 0.1007 | 0.2064 |
| Temperature | 0.5028 | 0.1749 | 0.0993 | 0.2064 |
| Platt | 0.4565 | 0.1529 | 0.0313 | 0.1055 |
| Generic | 0.4341 | 0.1396 | 0.0531 | 0.0910 |
| Generic+noise | 0.4555 | 0.1477 | 0.0360 | 0.1057 |
| Community | 0.4589 | 0.1439 | 0.0365 | 0.1102 |
| Combined | 0.4628 | 0.1359 | 0.0514 | 0.1104 |
| Guarded | 0.4313 | 0.1323 | 0.0489 | 0.0941 |

## IV. Experiments and Analysis

We compare known-type and held-out-type calibration, quantify the nuisance control gap, and test whether support-based routing recovers more loss than random fallback.

### A. Known-Type Calibration Before Shift

Table II reports all eight methods on the 800-sample known-type test view. Generic improves NLL from 0.5030 for uncalibrated predictions to 0.4341, while combined reaches 0.4628. Combined minus generic is +0.0287 NLL [0.0048, 0.0555], −0.0037 Brier [−0.0112, 0.0020], and +0.0194 AURC [0.0077, 0.0310]. Additional structure already has a mixed known-type tradeoff; the larger held-out-type NLL gap reflects a widening of this baseline difference.

### B. Do Community Covariates Improve Calibration?

Table III reports the predeclared combined-minus-generic comparison. Positive differences indicate worse performance after adding community covariates. Mean NLL increases by 0.1369; the type-cluster interval [−0.0125, 0.2864] includes zero. Combined improves in one of four held-out types and 3 of 12 fold-seed cells. Mean Brier score and AURC also worsen, with intervals that include zero.

The community-only calibrator also increases AURC by 0.0536 relative to the generic calibrator, with type-cluster interval [−0.0129, 0.1201]. Equal feature counts remove one simple capacity difference, but four held-out types leave substantial uncertainty.

### C. Does Extra Capacity Explain the Gap?

Combined has eight more inputs than generic. The capacity-matched control adds eight label-independent nuisance variables to generic and averages 20 deterministic realizations. Relative to generic, it increases mean NLL by 0.0415 with type-cluster interval [0.0314, 0.0527], Brier score by 0.0092 [0.0066, 0.0131], and AURC by 0.0039 [−0.0040, 0.0122]. Extra dimensions therefore impose a measurable finite-sample cost. The direct combined-minus-nuisance NLL difference is

TABLE III. PAIRED COMBINED-MINUS-GENERIC EFFECTS ON THE HELD-OUT-TYPE TEST VIEW. INTERVALS RESAMPLE FOUR HELD-OUT TYPES AFTER AVERAGING SEEDS.

| Metric | Mean diff. | 95% interval | Better types |
|---|---|---|---|
| NLL | +0.1369 | [ −0.0125, +0.2864 ] | 1/4 |
| Brier | +0.0189 | [ −0.0142, +0.0519 ] | 2/4 |
| AURC | +0.0362 | [ −0.0234, +0.0994 ] | 1/4 |

+0.0954 [−0.0637, 0.2458], with Brier +0.0097 [−0.0208, 0.0401] and AURC +0.0324 [−0.0199, 0.0849]. These intervals include zero. The smaller nuisance mean does not isolate support shift from feature correlation or finite-sample overfitting.

### D. *Can a Calibration-Support Guard Reduce the Loss?*

At the 95th-percentile threshold, the guard activates on 20.75% of the balanced shift view, ranging from 2.25% for addisplay to 52.25% for downloader. Relative to combined, it reduces NLL by 0.1133 [−0.1795, −0.0470], Brier by 0.0168 [−0.0287, −0.0061], and AURC by 0.0418 [−0.0747, −0.0114]. Across 90th–99th-percentile thresholds, all guarded-minus-combined NLL, Brier, and AURC intervals remain below zero.

Guarded NLL is 0.9522, compared with 1.0655 for combined and 0.9286 for generic. Its combined-minus-generic gap is reduced from +0.1369 to +0.0237, whose interval [−0.0626, 0.1069] includes zero. The guard therefore recovers most of the structural-calibration loss; it does not establish superiority over the generic baseline or the uncalibrated detector.

Matched-rate random fallback has mean NLL 1.0347. Guarded minus random fallback is −0.0825 [−0.1380, −0.0270] NLL, −0.0144 [−0.0217, −0.0070] Brier, and −0.0350 [−0.0611, −0.0088] AURC. The support score selects predictions with greater recoverable loss than random masks on these folds. Both the rule and this control were developed after the primary result, so the recovery remains exploratory.

### E. *Do Sample Size and Regularization Explain the Remaining Gap?*

We refit generic and combined calibrators from every retained checkpoint. Across $C \in \{0.01, 0.1, 1, 10\}$, mean combined-minus-generic NLL remains positive, ranging from +0.0209 to +0.1615. The type-cluster interval excludes zero at $C = 0.01$ and 0.1, but includes zero at $C = 1$ and 10. We also draw 10 deterministic, class-stratified calibration subsets per cell at sizes 40, 80, 120, and 160, and use the full 200 graphs as the endpoint. Mean NLL gaps decrease from +0.2775 at 40 graphs to +0.1369 at 200. Finite calibration data therefore amplify the degradation, while stronger regularization and the available sample-size range do not reverse its mean direction.

### F. *How Does the Effect Vary by Malware Type?*

Table IV separates the paired effects by held-out type. The sign of the NLL difference is stable across all three seeds within every type: combined improves on addisplay but worsens on adware, downloader, and trojan. Trojan has the largest NLL and AURC increases and twice as many high-confidence errors under combined as under generic. Downloader exhibits the largest univariate support shift, but not the largest loss increase.

TABLE V. MEAN PERFORMANCE ON THE 400-SAMPLE HELD-OUT-TYPE TEST VIEW. LOWER IS BETTER FOR NLL, BRIER, AND AURC; HIGHER IS BETTER FOR ACCURACY.

| Method | NLL | Brier | AURC | Acc. |
|---|---|---|---|---|
| Uncalibrated | **0.7632** | **0.2668** | **0.4414** | **0.5767** |
| Temperature | 0.7890 | 0.2703 | **0.4414** | **0.5767** |
| Platt | 0.8148 | 0.2961 | 0.4867 | 0.4852 |
| Generic | 0.9286 | 0.3170 | 0.4800 | 0.4592 |
| Generic+noise | 0.9701 | 0.3262 | 0.4839 | 0.4729 |
| Community | 0.9499 | 0.3284 | 0.5336 | 0.4325 |
| Combined | 1.0655 | 0.3359 | 0.5163 | 0.4838 |
| Guarded | 0.9522 | 0.3191 | 0.4745 | 0.5013 |

Thus, the aggregate result is not caused by one optimization seed, and displacement in a single covariate is not a calibrated predictor of degradation.

### G. *Absolute Performance and Detector Context*

Table V places the structural comparisons beside the frozen detector. The uncalibrated model remains best on all four columns, but its shift accuracy is only 0.5767 and its error-detection area under the receiver operating characteristic curve (AUROC) is 0.4902. The experiment therefore studies calibration during substantial detector degradation rather than a high-accuracy deployment regime. The nuisance control lies between generic and combined on NLL and Brier score. Guarded calibration recovers most of the combined loss and raises accuracy to 0.5013.

Fig. 1 shows how the guard changes predictions. Combined is overconfident across much of the probability range, whereas guarded probabilities move toward generic reliability and reduce risk over most coverage levels. A high-confidence error is an incorrect prediction with maximum class probability at least 0.9. Guarded produces 33.25 such errors per cell, compared with 43.25 for combined and 32.25 for generic.

### H. *Does Malware Prevalence Explain the Result?*

The known-type and primary held-out-type views differ in malware prevalence, so their direct contrast mixes type and class-prior shift. On the approximately prevalence-matched view, combined minus generic remains positive: 0.1055 NLL, 0.0285 Brier, and 0.0750 AURC. Guarded NLL is 0.9246, compared with 1.0137 for combined and 0.9081 for generic, and its high-confidence-error count falls to 23.67 from 31.83 for combined. Matching the class prior therefore preserves both the failure direction and most of the guard's recovery.

### I. *Where Does Degradation Occur?*

The loss gap varies with graph size. In the smallest graph-size quartile, combined NLL is 1.4523 versus 1.0767 for generic; in the largest quartile, it is 1.0662 versus 0.9562. Some held-out covariates also lie far from the known-type calibration distribution. In the downloader fold, absolute standardized mean differences reach 4.716 for component ratio and 4.622 for community-count ratio.

TABLE IV. FOLD-LEVEL DIAGNOSTICS, AVERAGED OVER THREE SEEDS. DIFFERENCES ARE COMBINED MINUS GENERIC; HCE IS THE NUMBER OF HIGH-CONFIDENCE ERRORS PER 400-SAMPLE CELL. THE LARGEST $|d|$ NEED NOT COINCIDE WITH THE LARGEST LOSS.

| Held-out type | ΔNLL | ΔBrier | ΔAURC | HCE gen. | HCE comb. | Max. $\|d\|$ | Feature |
|---|---|---|---|---|---|---|---|
| Addisplay | −0.1025 | −0.0210 | −0.0465 | 79.33 | 54.67 | 1.435 | Modularity |
| Adware | +0.2690 | +0.0567 | +0.0461 | 2.67 | 20.33 | 1.628 | Degree std. |
| Downloader | +0.0775 | −0.0073 | +0.0193 | 8.33 | 19.33 | 4.716 | Component ratio |
| Trojan | +0.3037 | +0.0471 | +0.1261 | 38.67 | 78.67 | 1.563 | Within-community density |

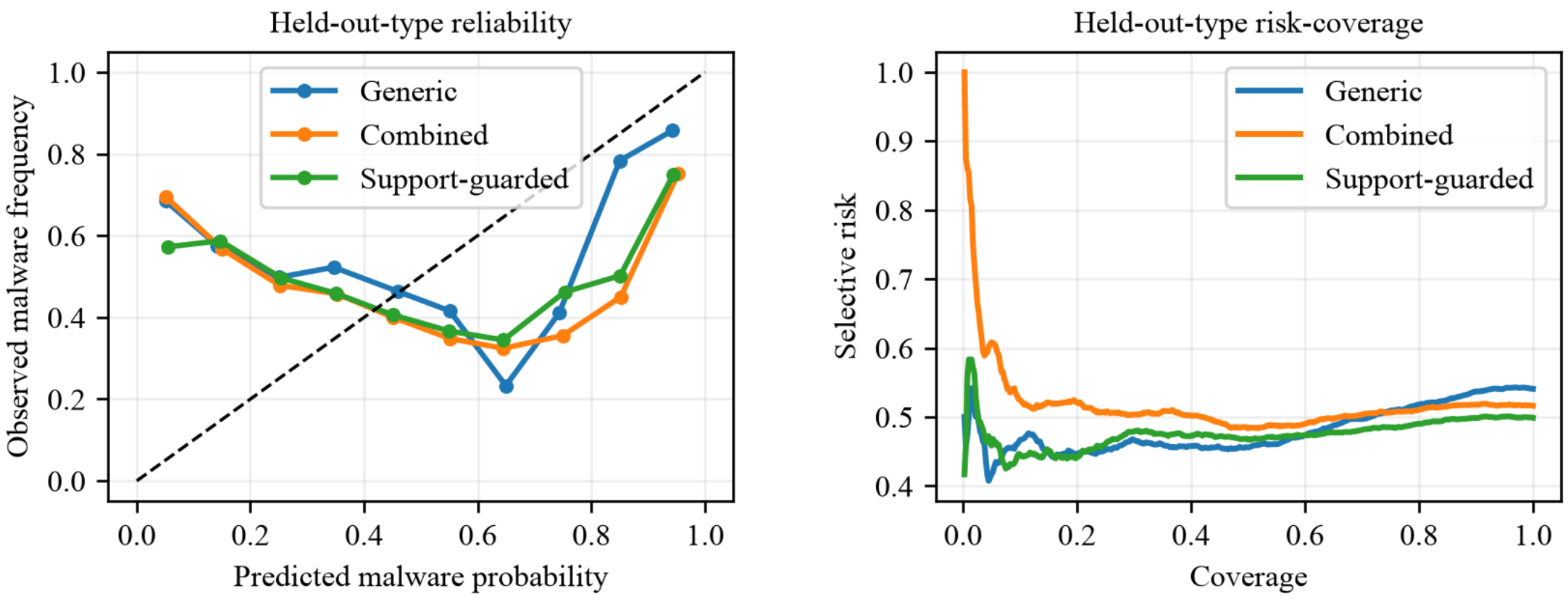

Fig. 1. Held-out-type reliability (left) and risk–coverage (right). The calibration-only support guard reduces combined overconfidence and selective risk while remaining close to the generic fallback.

These observations show that the held-out-type view contains structural regimes poorly represented during calibration and that degradation varies with graph size. They are consistent with support-sensitive extrapolation, but do not establish a causal feature or a monotonic relation between shift magnitude and performance. Graph size, components, community count, and the base logit can move together.

## V. Discussion and Limitations

### A. Operational Interpretation

Known-type tests show that generic calibration improves probability quality, while combined already increases NLL. Under type shift, uncalibrated predictions outperform every fitted mapping on NLL and Brier score. The support guard recovers combined loss and beats matched-rate random fallback on the same folds, but simpler alternatives remain stronger in absolute NLL. This is evidence of exploratory recovery, rather than an established deployment improvement. Downloader has the largest $|d|$, while trojan has the largest NLL increase; support distance is not a calibrated predictor of loss.

### B. Calibration and Decision Metrics

Classification records whether a probability crosses one decision boundary; NLL and Brier score evaluate its magnitude over the full interval. Combined can improve thresholded labels while assigning excessive confidence to remaining errors. Aggregate ECE also masks this behavior: combined has slightly lower fixed-bin ECE than generic but worse NLL, Brier score, AURC, and high-confidence-error count. Proper scores and full risk–coverage curves are therefore needed when probabilities drive ranking, abstention, or downstream review. At the system level, Xiao et al.'s CORA uses domain-constrained prompts and configuration and execution feedback for large language model (LLM)-driven OpenROADM automation [20].

### C. Reproducibility and Evidence Trace

The private companion repository archives experiment and analysis code, configuration, feature definitions, aggregate metrics, paired differences, and bootstrap summaries[1]. Per-sample outputs, checkpoints, and dataset caches remain locally retained. The base aggregate has 12 cells, 216 metric rows, and 100,800 predictions. Revision analyses verify stored guard probabilities against routing decisions and reproduce their NLL, Brier, and AURC before evaluating random masks. The addisplay run that reached the 80-epoch cap remains unchanged.

### D. Validity Boundaries

The conclusion is specific to MalNet-Tiny, a degree-feature GIN, deterministic Louvain summaries, logistic post-hoc models, and four held-out types. Each MalNet-Tiny malware type contains one family, so type and family effects cannot be separated. The shift is not temporal and does not measure zero-day or future-malware detection. Prevalence matching controls class prior but not every distributional change. Standardized shifts and graph-size strata remain observational diagnostics.

The calibration partition contains 200 graphs, while generic and community use nine inputs each and combined uses 17. Calibration-size refits show larger NLL gaps for smaller fitting sets, while regularization sweeps retain positive mean gaps throughout. The nuisance control shows a cost from additional dimensions; the uncertain direct comparison leaves community content, correlation, and support shift unresolved. Capacity, guard, and random-fallback analyses use the original folds after observing the primary result. Fresh malware types, another dataset, and stronger graph detectors are needed to test generalization. Three seeds and 10,000 bootstrap resamples quantify variability without increasing the four independent held-out-type units.

[1] https://github.com/junru-zhu/stress-testing-structure-aware-calibration

## VI. Conclusion

On MalNet-Tiny, combined calibration has higher known-type NLL than generic and a larger, heterogeneous held-out-type gap. Primary intervals and the combined-versus-nuisance interval include zero, leaving general average harm and its mechanism unresolved. A calibration-only support guard recovers combined loss and outperforms matched-rate random fallback on the original folds, while generic and uncalibrated predictions retain lower NLL. The stress test demonstrates why known-type baselines, paired controls, and explicit support diagnostics matter when assessing structural calibration.

## References

[1] P. Xu, C. Eckert, and A. Zarras, "Detecting and categorizing android malware with graph neural networks," in *Proceedings of the 36th Annual ACM Symposium on Applied Computing*, 2021, pp. 409–412.

[2] X. Ling, L. Wu, W. Deng, Z. Qu, J. Zhang, S. Zhang, T. Ma, B. Wang, C. Wu, and S. Ji, "MalGraph: Hierarchical graph neural networks for robust windows malware detection," in *IEEE INFOCOM 2022*, 2022, pp. 1998–2007.

[3] S. Freitas, Y. Dong, J. Neil, and D. H. Chau, "A large-scale database for graph representation learning," in *Advances in Neural Information Processing Systems, Datasets and Benchmarks Track*, 2021. [Online]. Available: https://openreview.net/forum?id=1xDTDk3XPW

[4] C. Guo, G. Pleiss, Y. Sun, and K. Q. Weinberger, "On calibration of modern neural networks," in *Proceedings of the 34th International Conference on Machine Learning*, vol. 70, 2017, pp. 1321–1330. [Online]. Available: https://proceedings.mlr.press/v70/guo17a.html

[5] F. Pendlebury, F. Pierazzi, R. Jordaney, J. Kinder, and L. Cavallaro, "TESSERACT: Eliminating experimental bias in malware classification across space and time," in *28th USENIX Security Symposium*, 2019, pp. 729–746. [Online]. Available: https://www.usenix.org/conference/usenixsecurity19/presentation/pendlebury

[6] D. Li, T. Qiu, S. Chen, Q. Li, and S. Xu, "Can we leverage predictive uncertainty to detect dataset shift and adversarial examples in android malware detection?" in *Annual Computer Security Applications Conference*, 2021, pp. 596–608.

[7] H. Li, G. Xu, L. Wang, X. Xiao, X. Luo, G. Xu, and H. Wang, "MalCertain: Enhancing deep neural network based android malware detection by tackling prediction uncertainty," in *Proceedings of the IEEE/ACM 46th International Conference on Software Engineering*, 2024, pp. 1–13.

[8] A. Herzog, A. Eusebi, and L. Cavallaro, "On the reliability and stability of selective methods in malware classification tasks," *arXiv preprint arXiv:2505.22843*, 2025. [Online]. Available: https://arxiv.org/abs/2505.22843

[9] M. A. Haque, I. Hossain, M. M. Kamol, M. J. Alam, S. K. Amalapuram, S. Talukder, and M. S. Rahman, "LAMDA: A longitudinal android malware benchmark for concept drift analysis," in *International Conference on Learning Representations*, 2026, poster. [Online]. Available: https://openreview.net/forum?id=1FnCrZtBNQ

[10] T. Bilot, N. El Madhoun, K. Al Agha, and A. Zouaoui, "A survey on malware detection with graph representation learning," *ACM Computing Surveys*, vol. 56, no. 11, pp. 1–36, 2024.

[11] L. Xiao, V. Gudepu, and A. Fumagalli, "Joint alarm correlation and root alarm detection in optical transport networks with graph-attention," in 2025 IEEE 30th International Workshop on Computer Aided Modeling and Design of Communication Links and Networks (CAMAD), 2025, pp. 1–6, doi: 10.1109/CAMAD67323.2025.11229899.

[12] L. Yang, W. Guo, Q. Hao, A. Ciptadi, A. Ahmadzadeh, X. Xing, and G. Wang, "CADE: Detecting and explaining concept drift samples for security applications," in *30th USENIX Security Symposium*, 2021. [Online]. Available: https://www.usenix.org/conference/usenixsecurity21/presentation/yang

[13] S. Park, O. Bastani, J. Weimer, and I. Lee, "Calibrated prediction with covariate shift via unsupervised domain adaptation," in *Proceedings of the Twenty Third International Conference on Artificial Intelligence and Statistics*, vol. 108, 2020, pp. 3219–3229. [Online]. Available: https://proceedings.mlr.press/v108/park20b.html

[14] A. Pampari and S. Ermon, "Unsupervised calibration under covariate shift," *arXiv preprint arXiv:2006.16405*, 2020. [Online]. Available: https://arxiv.org/abs/2006.16405

[15] X. Wang, H. Liu, C. Shi, and C. Yang, "Be confident! Towards trustworthy graph neural networks via confidence calibration," in Advances in Neural Information Processing Systems, vol. 34, 2021, pp. 23768–23779.

[16] A. Bahi, "When does distribution shift break graph neural networks calibration?" *arXiv preprint arXiv:2607.10804*, 2026. [Online]. Available: https://arxiv.org/abs/2607.10804

[17] K. Xu, W. Hu, J. Leskovec, and S. Jegelka, "How powerful are graph neural networks?" in *International Conference on Learning Representations*, 2019. [Online]. Available: https://openreview.net/forum?id=ryGs6iA5Km

[18] V. D. Blondel, J.-L. Guillaume, R. Lambiotte, and E. Lefebvre, "Fast unfolding of communities in large networks," *Journal of Statistical Mechanics: Theory and Experiment*, vol. 2008, no. 10, p. P10008, 2008.

[19] B. Efron, "Bootstrap methods: Another look at the jackknife," *The Annals of Statistics*, vol. 7, no. 1, pp. 1–26, 1979.

[20] L. Xiao, V. Gudepu, and A. Fumagalli, "CORA: A conversational AI-agent for intelligent automation of OpenROADM multi-vendor optical transport networks," in Applications of Machine Learning 2026, Proc. SPIE, vol. 14218, 2026, doi: 10.1117/12.3113861.

[21] Z. Liu, H. Lyu, J. Zhang, and Y. Xie, "LWCal: Loss-weighted calibration for tabular classifiers with noisy calibration labels," 2026. [Online]. Available: https://arxiv.org/abs/2609.26839

[22] Z. Liu, H. Lyu, J. Zhang, Y. Xie, and Q. Chen, "BenchGuard: A calibrated, adversary-aware triage layer for reward-hacking agentic trajectories," in *Proc. 35th ACM Int. Conf. Inf. Knowl. Manag.*, ser. CIKM '26. ACM, 2026.

[23] H. Yu, L. Liu, and D. Zhang, "Cost-aware post-hoc deferral under calibration and shift: An environmental AI case study," arXiv preprint arXiv:2609.09235, 2026. [Online]. Available: https://arxiv.org/abs/2609.09235

[24] Y. Zhang, A. Mehra, S. Niu, and J. Hamm, "DPCore: Dynamic prompt coreset for continual test-time adaptation," in Proceedings of the 42nd International Conference on Machine Learning, 2025, pp. 75757–75778. [Online]. Available: https://dl.acm.org/doi/abs/10.5555/3780338.3783377

[25] Y. Zhang, S. Niu, C. Cai, F. Liu, and J. Hamm, "Adapting in the dark: Efficient and stable test-time adaptation for black-box models," arXiv preprint arXiv:2604.15609, 2026. [Online]. Available: https://arxiv.org/abs/2604.15609